\documentclass[aps,preprint]{revtex4}%
\usepackage{amsfonts}
\usepackage{amsmath}
\usepackage{amssymb}
\usepackage{graphicx}%
\providecommand{\U}[1]{\protect\rule{.1in}{.1in}}

\begin{document}
\title[ ]{Concerning Classical Forces, Energies, and Potentials for Accelerated Point
Charges }
\author{Timothy H. Boyer}
\affiliation{Department of Physics, City College of the City University of New York, New
York, New York 10031, USA}

\begin{abstract}
Although the expressions for energy densities involving electric and magnetic
fields are exactly analogous, the connections to forces and electromagnetic
potentials are vastly different. \ For electrostatic situations, the changes
in the \textit{electric} energy can be related directly to \textit{electric}
forces and to the electrostatic potential$.$ \ The situation involving
magnetic forces and energy changes involves two fundamentally different
situations. \ For charged particles moving with constant velocities, the
changes in both electric and magnetic field energies are provided by the
external forces that keep the particles' velocities constant; there are no
Faraday acceleration electric fields in this situation. \ However, for
particles that change speed, the changes in \textit{magnetic} energy density
are related to acceleration-dependent Faraday \textit{electric} fields.
\ Current undergraduate and graduate textbooks deal only with highly symmetric
situations where the back\ Faraday electric fields are easily calculated from
the time-changing magnetic flux. \ However, in situations that lack high
symmetry, such as the magnetic Aharonov-Bohm situation, the back (Faraday)
acceleration electric fields of point charges seem unfamiliar. In this
article, we present a simple unsymmetric example and analyze it using the
Darwin Lagrangian. \ In \textit{all} cases involving changing velocities of
the current carriers, it is the work done by the back (Faraday) acceleration
\textit{electric} fields that balances the \textit{magnetic} energy changes.

\end{abstract}
\maketitle

\section{Introduction}

The expression $\mathbf{E}^{2}/\left(  8\pi\right)  $ for the energy density
associated with electric fields is exactly analogous to the energy density
$\mathbf{B}^{2}/\left(  8\pi\right)  $ for magnetic fields. \ However, the
connections of the energy expressions with the forces exerted by the fields
are vastly different between the electric and magnetic cases. \ The work done
by electrostatic forces leads directly to changes in electric field energy.
\ On the other hand, magnetic fields do no work. \ Accordingly, the
association between changes in magnetic energy and work are more subtle. \ In
the quasistatic regime, \textit{electric} field energies depend upon only the
relative \textit{positions} of charges, whereas \textit{magnetic} field
energies depend upon both the relative \textit{positions} of charges and also
the \textit{velocities} of the charges. \ Current textbooks of
electromagnetism discuss quasistatic magnetic energy changes for only two
situations: 1) charges that move with constant speed, and 2) charges which
change their speed but are in highly symmetric configurations. \ This limited
perspective leaves out magnetic energy changes for situations that lack high
symmetry. Here we present a simple point-charge example that lacks high
symmetry and illustrates the connections between forces, energies, and
potentials in quasistatic classical electrodynamics when radiation fields are
excluded. \ 

Although electrostatics and magnetostatics can be treated accurately using
continuous charge densities and currents, classical electrodynamics is a
relativistic theory and requires the use of point charges. \ Accordingly, we
will use point charges in our analysis of quasistatic energy changes. \ In
order to simplify the situation, we turn to the Darwin Lagrangian which treats
classical electrodynamics accurately through order $1/c^{2}$. \ Use of this
approximation simplifies the analysis by eliminating radiation fields and
retarded times. \ 

Reference to an analogy may clarify the purpose of the present article.
\ Magnetostatics, like electrostatics, involves only two of Maxwell's
equations for the field, $\nabla\cdot\mathbf{B}=0$ and $\nabla\times
\mathbf{B}=(4\pi/c)\mathbf{J}.$ \ However, if one considers only situations of
high symmetry, then one can apply only Ampere's law and ignore the divergence
requirement for $\mathbf{B}$. \ Thus if a current density $\mathbf{J}\left(
r\right)  $ has axial symmetry, then Ampere's law plus symmetry is sufficient
to determine the magnetic field $\mathbf{B}$, without any need for the
divergence equation. \ However, the highly-symmetric situations give a limited
and false impression of the theory. \ When the situation lacks high symmetry,
then the Biot-Savart law must be used. \ The Biot-Savart law is a particular
solution requiring \textit{both} Maxwell equations for $\mathbf{B}$, and may
be complemented by a solution of the homogeneous equations. \ Similarly,
highly-symmetric situations involving axial symmetry give the impression that
Faraday's law $\nabla\times\mathbf{E}=-\left(  1/c\right)  \partial
\mathbf{B}/\partial t$ can be used without the need for the rest of Maxwell's
equations because the confounding terms are suppressed by the symmetry.
\ However, the symmetric situations give only a limited understanding of
slowly-varying electromagnetic fields. \ 

This understanding limited to symmetric situations makes physicists unprepared
to analyze situations which lack high symmetry, such as the simple example in
the present article or such as the Aharonov-Bohm situation. \ Although the
azimuthally-symmetric vector potential associated with a circular solenoid is
often treated in junior-level courses in classical
electromagnetism,\cite{GriffithsA} the experimentally-realized situation of
the magnetic Aharonov-Bohm effect (where electrons pass on both sides of a
long solenoid) involves time-dependent interactions whose classical
electromagnetic aspects are \textit{not} azimuthally symmetric. \ It is often
claimed in electromagnetism texts\cite{Shadowitz} and repeatedly in quantum
texts\cite{quantum} that there is no classical electromagnetic interaction of
the solenoid back on the passing electrons. \ This claim is made despite the
magnetic energy changes associated with the overlap of the magnetic field of a
passing electron with the magnetic field of the solenoid, which changes depend
on the side of the solenoid on which the electron passes. \ However, such
no-interaction claims do not take into account the Faraday acceleration fields
of the charges in the solenoid which are related to the magnetic energy
changes. \ Faraday acceleration fields associated with magnetic energy changes
are treated in the present article, but discussion of the Aharonov-Bohm
situation is made elsewhere.\cite{AB} \ \ \ 

\section{The Darwin-Lagrangian Approximation}

\subsubsection{The Darwin Lagrangian}

The classical electrodynamics of charged particles at arbitrary changing
velocities is enormously complicated, particularly because electromagnetic
fields depend upon sources at retarded times. \ In order to simplify the
situation, we will consider the behavior of just two point charges that move
at speeds small compared to the speed $c$ of light in vacuum. \ This situation
corresponds to that described by the Darwin Lagrangian\cite{Darwin}%
\cite{Jackson593}. This approximation represents an enormous simplification;
it excludes radiation and involves no retarded times.

The Darwin Lagrangian for two point charges $e$ and $q$ is given
by\cite{Darwin}\cite{Jackson593}%
\begin{align}
L  &  =-m_{e}c^{2}\sqrt{1-v_{e}^{2}/c^{2}}-m_{q}c^{2}\sqrt{1-v_{q}^{2}/c^{2}%
}-\frac{eq}{\left\vert \mathbf{r}_{e}\mathbf{-r}_{q}\right\vert }+\frac
{eq}{2c^{2}}\left[  \frac{\mathbf{v}_{e}\cdot\mathbf{v}_{q}}{\left\vert
\mathbf{r}_{e}\mathbf{-r}_{q}\right\vert }+\frac{\left[  \mathbf{v}%
_{e}\mathbf{\cdot}\left(  \mathbf{r}_{e}\mathbf{-r}_{q}\right)  \right]
\left[  \mathbf{v}_{q}\cdot\left(  \mathbf{r}_{e}\mathbf{-r}_{q}\right)
\right]  }{\left\vert \mathbf{r}_{e}\mathbf{-r}_{q}\right\vert ^{3}}\right]
\nonumber\\
&  =-m_{e}c^{2}\sqrt{1-v_{e}^{2}/c^{2}}-m_{q}c^{2}\sqrt{1-v_{q}^{2}/c^{2}%
}-\frac{1}{2}\left[  e\Phi_{q}(\mathbf{r}_{e},t)+q\Phi_{e}\left(
\mathbf{r}_{q},t\right)  \right] \nonumber\\
&  +\frac{1}{2}\left[  e\frac{\mathbf{v}_{e}}{c}\cdot\left[  \mathbf{A}%
_{q}(\mathbf{r}_{e},t)\right]  +q\frac{\mathbf{v}_{q}}{c}\cdot\left[
\mathbf{A}_{e}\left(  \mathbf{r}_{q},t\right)  \right]  \right]  \label{Lagra}%
\end{align}
where it is understood that the square roots are to be expanded in powers of
$1/c^{2}$ through order $1/c^{4}$. \ \ Here we have not expanded the square
roots because maintaining the exact mechanical forms leads to the familiar
expressions for the particle mechanical energies on calculations from the
Lagrangian. \ 

\subsection{Potentials and Fields in the Darwin-Lagrangian Approximation}

For a point charge $q$ located at $\mathbf{r}_{q}\left(  t\right)  $ and
moving with small velocity $\mathbf{v}_{q}\left(  t\right)  $ and acceleration
$\mathbf{a}_{q}\left(  t\right)  $, through order $1/c^{2}$, the system
Lagrangian in Eq.(\ref{Lagra}) can be described as involving a scalar
potential
\begin{equation}
\Phi_{q}(\mathbf{r,}t)=\frac{q}{\left\vert \mathbf{r-r}_{q}\left(  t\right)
\right\vert }, \label{PHq}%
\end{equation}
a vector potential\cite{Jackson593}%
\begin{equation}
\mathbf{A}_{q}\left(  \mathbf{r,}t\right)  =\frac{q}{2c}\left[  \frac
{\mathbf{v}_{q}}{\left\vert \mathbf{r-r}_{q}\right\vert }+\frac{\left[
\mathbf{v}_{q}\mathbf{\cdot}\left(  \mathbf{r-r}_{q}\right)  \right]  \left(
\mathbf{r-r}_{q}\right)  }{\left\vert \mathbf{r-r}_{q}\right\vert ^{3}%
}\right]  \mathbf{,} \label{Aq}%
\end{equation}
an electric field\cite{PA}%
\begin{align}
\mathbf{E}_{q}\mathbf{(r},t)  &  =-\nabla\Phi_{q}\left(  \mathbf{r,}t\right)
-\frac{1}{c}\frac{\partial\mathbf{A}_{q}\left(  \mathbf{r,}t\right)
}{\partial t}\nonumber\\
&  =q\frac{\mathbf{r-r}_{q}}{\left\vert \mathbf{r-r}_{q}\right\vert ^{3}%
}\left[  1+\frac{v_{q}^{2}}{2c^{2}}-\frac{3}{2}\left(  \frac{\mathbf{v}_{q}%
}{c}\mathbf{\cdot}\frac{\mathbf{r-r}_{q}}{\left\vert \mathbf{r-r}%
_{q}\right\vert }\right)  ^{2}\right] \nonumber\\
&  -\frac{q}{2c^{2}}\left[  \frac{\mathbf{a}_{q}}{\left\vert \mathbf{r-r}%
_{q}\right\vert }+\frac{\left[  \mathbf{a}_{q}\cdot\left(  \mathbf{r-r}%
_{q}\right)  \right]  \left(  \mathbf{r-r}_{q}\right)  }{\left\vert
\mathbf{r-r}_{q}\right\vert ^{3}}\right]  , \label{Eqrt}%
\end{align}
and a magnetic field
\begin{equation}
\mathbf{B}_{q}\left(  \mathbf{r,}t\right)  =\nabla\times\mathbf{A}_{q}\left(
\mathbf{r},t\right)  =q\frac{\mathbf{v}_{q}\left(  t\right)  }{c}\times
\frac{\mathbf{r-r}_{q}\left(  t\right)  }{\left\vert \mathbf{r-r}_{q}\left(
t\right)  \right\vert ^{3}}. \label{Bq}%
\end{equation}

\subsection{Comments on the Electromagnetic Fields in the Darwin
Approximation}

We notice immediately that the scalar potential in Eq. (\ref{PHq}) is the same
as that for a point charge in electrostatics, except that the position
$\mathbf{r}_{q}(t)$ is now time dependent. \ Just how different the theory is
from electrostatics can be seen in the expression (\ref{Eqrt}) for the
electric field which now involves velocities and even accelerations of the
charged particle. \ Similarly, the magnetic field in Eq. (\ref{Bq}) is that
which one might be tempted to use for a point charge from the Biot-Savart law
in magnetostatics using $\mathbf{J}_{q}\left(  \mathbf{r,}t\right)
=q\mathbf{v}_{q}\left(  t\right)  \delta^{3}\left(  \mathbf{r-r}_{q}\left(
t\right)  \right)  $. \ However, these expressions are now only approximations
since we are not dealing with electrostatics or magnetostatics. The scalar
potential appears in the connection with the electric field $\mathbf{E}%
_{q}\mathbf{(r},t)$ in Eq. (\ref{Eqrt}), but not in connection with the
magnetic field $\mathbf{B}_{q}\left(  \mathbf{r},t\right)  $ in Eq.
(\ref{Bq}). \ The scalar potential $\Phi_{q}(\mathbf{r,}t)$ depends on the
relative positions $\left\vert \mathbf{r-r}_{q}\right\vert $\ of the charge
and the field point. On the other hand, the vector potential $\mathbf{A}%
_{q}\left(  \mathbf{r,}t\right)  $ is connected to \textit{both}
$\mathbf{E}_{q}\mathbf{(r},t)$ and $\mathbf{B}_{q}\left(  \mathbf{r},t\right)
$. \ The vector potential $\mathbf{A}_{q}\left(  \mathbf{r,}t\right)  $
depends not only on the relative displacement $\left(  \mathbf{r-r}%
_{q}\right)  $ between the charge and the field point, but also on the
velocity $\mathbf{v}_{q}$ of the charge $q$. \ The magnetic field
$\mathbf{B}_{q}\left(  \mathbf{r,}t\right)  $ depends upon \textit{spatial}
derivatives of $\mathbf{A}_{q}\left(  \mathbf{r,}t\right)  $. \ In contrast,
the electric field $\mathbf{E}_{q}\mathbf{(r},t)$ in Eq. (\ref{Eqrt}) depends
upon the \textit{time }rate of change of the vector potential $\mathbf{A}%
_{q}\left(  \mathbf{r,}t\right)  $, which may involve changes in both the
charge's position $\mathbf{r}_{q}$ and/or the charge's velocity $\mathbf{v}%
_{q}$. \ The terms in Eq. (\ref{Eqrt}) for $\mathbf{E}_{q}\mathbf{(r},t)$
involving $v_{q}^{2}$ arise from \textit{position}-dependent changes in
$\mathbf{A}_{q}\left(  \mathbf{r,}t\right)  $ with time. \ The terms involving
acceleration $\mathbf{a}_{q}$ arise from time-dependent changes associated
with \textit{velocity}. The acceleration-dependent terms give the Faraday
acceleration fields of the charged particle. \ The acceleration-dependent
terms for the electric field in Eq. (\ref{Eqrt}) appear in an electromagnetism
textbook\cite{PA} published in 1940, but do not seem to appear in more recent textbooks.\ \ 

\subsection{Equations of Motion from the Darwin Lagrangian}

The Euler-Lagrange equations of motion for the charges $e$ and $q$ can be
obtained in terms of the canonical momentum of a particle. \ \ However, these
equations can be rewritten in the conventional form for Newton's second law as
$d\mathbf{p}_{mechanical}/dt=q\mathbf{E+}q(\mathbf{v}/c)\times\mathbf{B}$
where $\mathbf{E=-\nabla}\Phi\mathbf{-}\left(  1/c\right)  \partial
\mathbf{A}/\partial t$ and $\mathbf{B=\nabla\times A}$. \ 

\subsection{The Total Energy}

The associated system total energy follows directly from the Lagrangian in Eq.
(\ref{Lagra}), giving a sum of mechanical, electric, and magnetic energies,
\begin{equation}
U^{\left(  total\right)  }=U^{\left(  m\right)  }+U^{(E)}+U^{\left(  B\right)
},\label{Utotal1a}%
\end{equation}
where the mechanical energy is%
\begin{equation}
U^{\left(  m\right)  }=\frac{m_{e}c^{2}}{\sqrt{1-v_{e}^{2}/c^{2}}}+\frac
{m_{q}c^{2}}{\sqrt{1-v_{q}^{2}/c^{2}}},
\end{equation}
the electric field energy is%
\begin{equation}
U^{(E)}=\frac{1}{2}\left[  e\Phi_{q}(\mathbf{r}_{e},t)+q\Phi_{e}\left(
\mathbf{r}_{q},t\right)  \right]  =\frac{eq}{\left\vert \mathbf{r}%
_{e}\mathbf{-r}_{q}\right\vert },\label{UE}%
\end{equation}
and the magnetic field energy is
\begin{align}
U^{\left(  B\right)  } &  =\frac{1}{2}\left[  e\frac{\mathbf{v}_{e}}{c}%
\cdot\left[  \mathbf{A}_{q}(\mathbf{r}_{e},t)\right]  +q\frac{\mathbf{v}_{q}%
}{c}\cdot\left[  \mathbf{A}_{e}\left(  \mathbf{r}_{q},t\right)  \right]
\right]  \nonumber\\
&  =\frac{eq}{2c^{2}}\left[  \frac{\mathbf{v}_{e}\cdot\mathbf{v}_{q}%
}{\left\vert \mathbf{r}_{e}\mathbf{-r}_{q}\right\vert }+\frac{\left[
\mathbf{v}_{e}\mathbf{\cdot}\left(  \mathbf{r}_{e}\mathbf{-r}_{q}\right)
\right]  \left[  \mathbf{v}_{q}\cdot\left(  \mathbf{r}_{e}\mathbf{-r}%
_{q}\right)  \right]  }{\left\vert \mathbf{r}_{e}\mathbf{-r}_{q}\right\vert
^{3}}\right]  .\label{UB}%
\end{align}
\ 

\section{Field-Potential-Energy Connections for Electric and Magnetic Fields}

\subsection{Electric Energy Changes}

In the Darwin approximation, the changes in electric field energy are directly
analogous to those familiar from electrostatics. \ The electric energy in Eq.
(\ref{UE}) involves only the scalar potentials which take the same from as in
electrostatics. \ The electric forces associated with the electric potential
are directly associated with energy in the electric field. \ The electric
energy involves only the separations $\left\vert \mathbf{r}_{e}\mathbf{-r}%
_{q}\right\vert $ between the charged particles. \ 

\subsection{Magnetic Energy Changes}

In contrast to electric field energy changes which involve only the
separations between the charges, magnetic energy changes involve both changes
in the separations between the charges and also changes in the velocities of
the charges, as indicated in Eq. (\ref{UB}). \ It seems convenient to separate
magnetic energy changes into two separate situations--those involving charged
particles moving with constant velocity and those involving accelerating
charges. \ 

\subsubsection{Charged Particles Moving with Constant Velocity}

The situation of charges moving with constant velocities turns out to be
uncomplicated, and the connections between forces and energy changes are
\textit{exactly} soluble. \ It can be shown\cite{B1971} that for point charges
moving with \textit{constant velocities}, the external forces that hold the
particles at constant velocities also provide \textit{exactly} the changing
energies in the electric and magnetic fields for any particle velocities less
than the speed of light in vacuum. \ There are no kinetic energy changes and
no Faraday acceleration electric fields. \ 

A charged particle at rest in an inertial frame does not cause any $emf$
around any closed curve in that space. \ However, a charge particle moving
with constant velocity will indeed cause an $emf$ around a general closed
curve in the inertial frame. \ Also, the charge moving with constant velocity
has an associated magnetic field, and the $emf$ caused by the moving charge
agrees with the changing magnetic flux through the closed curve. \ Thus,
depending upon the speed of the charge in an inertial frame, a charge $q$ may
or may not cause an $emf$. Indeed, if we are dealing with only two charges
moving with constant velocities, we can always go to the rest frame of one of
the charges and so have changes in only electric field energy and no magnetic
field energy changes at all. \ 

Situations involving \textit{motional emfs} are often treated as though the
relevant charge carriers were moving with constant velocity, and so can be
understood in terms of energy changes being balanced by the work done by
external forces\cite{Griffiths}. \ 

\subsubsection{Accelerating Charges}

The particularly-troubling magnetic energy situations involve groups of
charges where particles change their speeds. \ These cases generally involve
Faraday acceleration electric forces, and the energy changes may involve
particle kinetic energy changes and/or changes in electromagnetic field energy
and/or work done by external forces. \ We will discuss magnetic energy changes
in connection with two different examples, one involving a highly symmetric
situation and one involving an unsymmetric situation. \ 

\section{ \ Situations Involving a Long Solenoid }

\subsection{Azimuthally Symmetric Solenoid with Increasing Currents}

The most familiar situation of a back (Faraday) acceleration electric field
occurs for the highly-symmetric situation of a circular solenoid. \ When the
currents in the solenoid are constant, the charge carriers are
\textit{accelerating} towards the central axis since they are moving in a
circle. \ A single charge moving in a circle does not have just electrostatic
and magnetostatic fields, but rather has higher terms in $1/c$, including
radiation terms. \ However, the situation changes for a many-particle current
which can be approximated as a steady current. \ Even though the charges are
still accelerating when the current in a circuit is constant, all the
higher-order contributions to the fields cancel, leaving only the
electrostatic and magnetostatic contributions\cite{J697}. \ For a circular
solenoid, the vector potential $\mathbf{A}$ inside may be chosen as symmetric
in the azimuthal $\widehat{\phi}$-direction with $\mathbf{A=\widehat{\phi}%
}Br/2\mathbf{=}\widehat{\phi}2\pi nevr/c$, and outside the solenoid
$\mathbf{A=}\widehat{\phi}2\pi nevR^{2}/\left(  cr\right)  $, where $n$ is the
number of current carriers per unit area, $e$ and $v$ are the charge and speed
of each current carrier, and $R$ is the radius of the long solenoid. \ This
corresponds to the Coulomb gauge for $\mathbf{A}$.

When the currents in a solenoid are changing in an azimuthally symmetric
pattern, the charge carriers in the winding of the solenoid have a tangential
acceleration associated   with the same change in \textit{speed }$v$ for each
charge. \ If there were no change in magnetic field energy, the agent causing
the changing speed of the charge carriers would need to provide power so as to
change only the \textit{kinetic energy} of the current carriers. \ This change
in kinetic energy of the current carriers is so small as to escape any mention
in textbooks of classical electromagnetism. \ It is the change in
\textit{magnetic field energy} that is overwhelmingly dominant. \ The agent
changing the speed of the current carriers must provide the power against the
back (Faraday) acceleration electric fields corresponding to the acceleration
terms in Eq. (\ref{Eqrt}). \ We emphasize: \textit{The magnetic energy balance
for quasistatic systems requires the existence of Faraday electric forces
associated with the accelerations of the charged particles.}

The back electric fields associated with the changing speeds of the charge
carriers can be described in terms of the changing vector potential. \ Thus,
for a long solenoid with slowly increasing currents, the \textit{changes of
speed} of all the current carriers give a changing vector potential with
non-zero $\partial\mathbf{A/\partial}t$ and so a back \textit{emf} from the
non-vanishing (Faraday) acceleration electric field appearing from
$\mathbf{E=-}\nabla\Phi\mathbf{-(}1/c\mathbf{)\partial A/\partial}t$.
\ Although the speed $v$ of all the particles changes (producing a changing
magnetic field and magnetic energy), the functional \textit{spatial}
dependence of the magnetic field $\mathbf{B}$ on the vector potential
$\mathbf{A}$ involves only \textit{spatial} separations and does not change,
$\mathbf{B=\nabla\times A}$. \ Inside the solenoid, an increasing vector
potential is associated with increasing speed $v$ and an increasing magnetic
field, since inside, $\mathbf{A}$ depends linearly on the distance $r$ from
the central axis $\mathbf{A=\widehat{\phi}}Br/2\mathbf{=}\widehat{\phi}2\pi
nevr/c$ and $\mathbf{B=\nabla\times A}\neq0.$ \ \ However, outside the
solenoid, even though the magnitude of the vector potential is increasing in
$\mathbf{A=}\widehat{\phi}2\pi nevR^{2}/\left(  cr\right)  $ due the increase
in $v$, the $1/r$ fall-off of the vector potential with distance from the
central axis still gives $\mathbf{B}=\nabla\times\mathbf{A}=0$. \ To the
surprise of many students, there is an acceleration-dependent
\textit{electric} field outside the solenoid from $\partial\mathbf{A/\partial
}t$, but still no \textit{magnetic} field outside. \ 

In our elementary electromagnetism classes where our examples and problems are
highly symmetric, it is easy to relate the back \textit{emf }of the solenoid
to Faraday's law involving a changing magnetic field,
\begin{equation}
emf=\oint d\mathbf{r}\cdot\mathbf{E}=-\frac{1}{c}\frac{d}{dt}\left(  \int
d\mathbf{S}\mathfrak{\cdot}\mathbf{B}\right)  \mathbf{,}%
\end{equation}
where $\mathbf{S}$ is the area, and to take advantage of the high symmetry to
calculated $\mathbf{E}$. \ The textbooks emphasize that changing currents in a
solenoid lead to changing magnetic fields and therefore to Faraday electric
fields. \ They do not emphasize that this back $emf$ is due directly to the
electric fields produced by the \textit{accelerating} current carriers.
\ There is no need to go through a changing magnetic field to obtain the
$emf$. \ The same current carriers that produce the Faraday acceleration
electric field also produce the magnetic field of the solenoid in the first
place\cite{self}. \ 

\subsection{Unsymmetric Situations}

The Darwin Lagrangian is rarely mentioned in junior-level electromagnetism
texts. \ Although the full Lienard-Wiechert fields involving retarded times
are often given, the quasistatic approximation through $1/c^{2}$ for the
electric fields in Eq. (\ref{Eqrt}) goes unmentioned even in graduate-level
texts. \ Thus many physicists are unfamiliar with the acceleration-dependent
terms for the electric field of a point charge that are given in Eq.
(\ref{Eqrt}), and so they are unable to understand the classical
electromagnetic aspects of an unsymmetric situation such as that proposed by
Aharonov-Bohm where electrons pass a long solenoid. \ 

When an electron passes alongside a long solenoid, the magnetic field of the
passing charge will overlap with the magnetic field of the solenoid giving a
change in the total magnetic energy, a relativistic $1/c^{2}$ effect.
\ Depending upon which side the electron passes, the change in total magnetic
field energy can be positive or negative. \ Naturally, one asks,
\textquotedblleft What provides the energy which balances the magnetic energy
change?\textquotedblright\ \ If the solenoid is a superconductor, there is no
external source of energy. \ In other cases, there is a battery providing
current to the solenoid. \ In this latter case, one sometimes hears the
suggestion that the battery provides the necessary energy changes. \ However,
this side-dependent magnetic energy change does \textit{not} correspond to a
highly symmetric situation. Indeed, a relativistic $1/c^{2}$ effect is
involved, and the forces on the charged particles of the solenoid due to the
fields of a passing charge will be in \textit{different directions in
different locations}, some solenoid charges increasing their speeds and some
decreasing their speeds. For situations that are not highly symmetric, we
cannot easily use the familiar two-step procedure through the magnetic field
to obtain the induced electric field. \ 

In the past, it has been suggested repeatedly\cite{lagg} that the full
magnetic energy change associated with a passing electron was balanced by the
kinetic energy change of the passing electron itself. \ Such a kinetic energy
change for a passing charge indeed accounts for the magnitude of the
experimentally observed Aharonov-Bohm phase shift in terms of a classical
electromagnetic lag effect rather than the currently prevailing idea of a
\textquotedblleft purely quantum topological effect\textquotedblright\ of
magnetic flux leading to the Aharonov-Bohm phase shift. \ In any case, the
classical electromagnetic situation of charged particles passing a long
solenoid lacks high symmetry, and therefore we cannot use symmetry and the
integral form of Faraday's law to calculate the force on the passing
electrons. \ One must turn to a more fundamental analysis, such as one through
the Darwin Lagrangian. \ 

\section{Example of Two Point Charges Moving Side-by-Side}

\subsection{Side-by-Side Charges}

For the simplest possible illustration of magnetic energy changes in a
situation lacking high symmetry, we consider two point charges $e$ and $q$
moving side-by-side at separation $l=\left\vert \mathbf{r}_{e}\mathbf{-r}%
_{q}\right\vert $ along frictionless rods parallel to the $z$-axis with
velocities $\mathbf{v}_{e}$ and $\mathbf{v}_{q}$. \ This system will have both
electric and magnetic fields and associated energies. \ For motion with
\textit{constant velocity}, the electric and magnetic forces of one charge on
the other are perpendicular to the frictionless rods. This same system, where
the forces of constraint introduce neither energy nor net linear momentum, was
used previously\cite{Emc2} as an illustration of Einstein's mass-energy
relation $\mathcal{E}=mc^{2}$.

In Darwin's $1/c^{2}$ approximation for the energy (given in Eqs.
(\ref{Utotal1a})-(\ref{UB})), the electric field energy is still $U^{\left(
E\right)  }(\mathbf{r}_{e},\mathbf{r}_{q})=eq/\left\vert \mathbf{r}%
_{e}\mathbf{-r}_{q}\right\vert $ in Eq. (\ref{UE}), whereas the magnetic field
energy is given in Eq. (\ref{UB}). \ For our simple side-by-side situation,
where the separation $l=\left\vert \mathbf{r}_{e}\mathbf{-r}_{q}\right\vert $
of the charges is perpendicular to the direction of motion $\mathbf{v}$, we
have the magnetic field energy%
\begin{equation}
U^{\left(  B\right)  }\left(  \mathbf{r}_{e},\mathbf{r}_{q},\mathbf{v}\right)
=\frac{1}{2}\left[  e\frac{v}{c}\left(  \frac{q}{2c}\frac{v}{l}\right)
+q\frac{v}{c}\left(  \frac{e}{2c}\frac{v}{l}\right)  \right]  =\frac{eqv^{2}%
}{2c^{2}l}. \label{UBeql}%
\end{equation}

\subsection{Accelerating the System of Two Point Charges}

Suppose that external forces $\mathbf{F}_{ext}^{\left(  e\right)  }$ and
$\mathbf{F}_{ext}^{\left(  q\right)  }$ parallel to the $z$-axis are applied
to the two charges so as to give the particles a small common acceleration
$\mathbf{a}=\widehat{z}a$\ for a short time $\tau$, thus changing the
particles' common speed by $\Delta v=a\tau$. \ The kinetic energy
$mc^{2}\left(  \gamma-1\right)  $ of each particle changes due to the change
in speed. \ The field energy in the electric field does not change according
to the Darwin approximation in Eq. (\ref{UE}). \ However, the energy in the
magnetic field changes from that given in Eq. (\ref{UBeql}) over to that with
$v$ replaced by $v+\Delta v$, so that the change in magnetic energy is%
\begin{equation}
\Delta U^{\left(  B\right)  }=\frac{eq\left(  v+\Delta v\right)  ^{2}}%
{2c^{2}l}-\frac{eqv^{2}}{2c^{2}l}\approxeq\frac{eqv\Delta v}{c^{2}l}.
\label{DUB1}%
\end{equation}

During the acceleration of the parallel-moving charges $e$ and $q$, the vector
potential $\mathbf{A=A}_{e}+\mathbf{A}_{q}$ changes as the speed changes,
leading to changes in the magnetic field $\mathbf{B=\nabla\times A}$ and in
the magnetic field energy. \ However, the acceleration of the charges and the
change in the vector potential $\mathbf{A}$ also leads to a change in the
electric field through $\mathbf{E=-\nabla}\Phi\mathbf{-}\left(  1/c\right)
\partial\mathbf{A}/\partial t$. Thus, the electric field has a new
\textit{acceleration-related} component \textit{parallel} to the velocity
$\mathbf{v}$ of the particles. \ Therefore, from Eq. (\ref{Eqrt}), each charge
exerts an additional (\textit{acceleration-dependent}) electric force
\begin{equation}
e\mathbf{E}_{acc-q}=q\mathbf{E}_{acc-e}=-\frac{eq\mathbf{a}}{2c^{2}\left\vert
\mathbf{r}_{e}\mathbf{-r}_{q}\right\vert }=-\frac{eq\mathbf{a}}{2c^{2}l},
\end{equation}
a \textit{retarding} force, on the other charge. \ But now Newton's second law
for the \textit{mechanical} momentum change in the $z$-direction for each
particle involves not just one external force on each particle, but rather two
forces, both the external force and the electric force of one charge on the
other,
\begin{equation}
\frac{d\mathbf{p}_{e}}{dt}=\mathbf{F}_{ext}^{\left(  e\right)  }%
-\frac{eq\mathbf{a}}{2c^{2}l}\text{ \ \ and \ \ }\frac{d\mathbf{p}_{q}}%
{dt}=\mathbf{F}_{ext}^{\left(  q\right)  }-\frac{qe\mathbf{a}}{2c^{2}%
l}.\label{Nsl}%
\end{equation}
\ Therefore, the external forces (which provided the acceleration $\mathbf{a}$
of the charges) have to be larger so as to overcome these retarding forces.
\ Rewriting Newton's second-law equations in (\ref{Nsl}) by moving the
electric forces to the opposite sides of the equal signs, we require%
\begin{equation}
\mathbf{F}_{ext}^{\left(  e\right)  }=\frac{d\mathbf{p}_{e}}{dt}%
+\frac{eq\mathbf{a}}{2c^{2}l}\text{ \ \ and \ \ }\mathbf{F}_{ext}^{\left(
q\right)  }=\frac{d\mathbf{p}_{q}}{dt}+\frac{qe\mathbf{a}}{2c^{2}l}.
\end{equation}
The change in the energy of the system due to the external forces is
\begin{align}
\Delta U^{\left(  total\right)  } &  =\int_{0}^{\tau}dt\left(  \mathbf{F}%
_{ext}^{\left(  e\right)  }+\mathbf{F}_{ext}^{\left(  q\right)  }\right)
\cdot\mathbf{v=}\int_{0}^{\tau}dt\left(  \frac{d\mathbf{p}_{e}}{dt}%
+\frac{d\mathbf{p}_{q}}{dt}+\frac{eq\mathbf{a}}{c^{2}l}\right)  \cdot
\mathbf{v}\nonumber\\
&  \approxeq\Delta U^{\left(  m_{e}\right)  }+\Delta U^{\left(  m_{q}\right)
}+\frac{eq\mathbf{v\cdot a}\tau}{c^{2}l}=\Delta U^{\left(  m_{e}\right)
}+\Delta U^{\left(  m_{q}\right)  }+\frac{eqv\Delta v}{c^{2}l},
\end{align}
where we have used $\Delta\mathbf{v=a}\tau$ and have kept only first-order
terms. \ Thus the external forces indeed account for the energy change, both
the \textit{mechanical} energy change $\Delta U^{\left(  m_{e}\right)
}+\Delta U^{\left(  m_{q}\right)  }$ (from the time-integrations over the
changes in momentum), and also the \textit{magnetic} energy change (from the
time-integration over the Faraday acceleration electric fields) corresponding
to Eq. (\ref{DUB1}).

In this example, the external forces had to be larger because of the changing
vector potential $\mathbf{A=A}_{q}+\mathbf{A}_{e}$, which gave a new
acceleration-dependent term in the expression $\mathbf{E=-\nabla}\Phi-\left(
1/c\right)  \partial\mathbf{A/\partial}t$. \ This is a common situation for
charges that are changing speed. \ \ \textit{The magnetic energy balance for
quasistatic systems requires the existence of Faraday electric forces
associated with the accelerations of the charged particles.}

Thus for unsymmetrical situations (such as the simple example here of charges
accelerating side-by-side, or indeed of charges passing outside a solenoid in
the Aharonov-Bohm situation), we can not use high symmetry together with
Faraday's law involving changing magnetic flux, but rather we must fall back
on more fundamental considerations; we need to connect changes in magnetic
field energy directly with particle accelerations and with the back (Faraday)
acceleration terms in Eq. (\ref{Eqrt}) for the electric field of a point
charge. \ Indeed, for groups of closely-spaced charges, the inertial effects
of the back (Faraday) acceleration field become important, and failure to
recognize these fields in unsymmetric cases can lead to paradoxes\cite{AB}.

This example involving two point charges illustrates some of the contrasts
involving forces, energies, and potentials for the electric and magnetic
fields. \ In the electric case, the work associate with the electric field
appears directly as energy stored in the electric fields, as is familiar in
electrostatics. \ In the magnetic case, external or internal forces can change
the speeds of the particles that are producing the magnetic fields. \ The
change in the speeds of the charged particles leads to changes in magnetic
energy. \ However, the change of charged particles' speeds also causes back
(Faraday) \textit{acceleration} electric fields which make it harder for the
external or internal forces to accelerate the charges. \ It is the additional
work against the back (Faraday) acceleration electric fields that balances the
change in magnetic field energy.

\section{Closing Summary}

The Darwin energy expressions in Eqs. (\ref{Utotal1a})-(\ref{UB}) give the
energy of two low-speed interacting charges. \ The electric field energy is
the familiar expression from electrostatics. \ The magnetic field energy
involves the separations and velocities of the charged particles. \ The
magnetic field energy can change while keeping the particle velocities
$\mathbf{v}_{e}$ and $\mathbf{v}_{q}$ constant if the relative displacement
$\left(  \mathbf{r}_{e}-\mathbf{r}_{q}\right)  $ between the charges changes.
\ In this case of charges moving with constant velocities, the change in
magnetic field energy is provided by the external forces which keep the
charges moving with constant velocity\cite{B1971}. \ There are no Faraday
acceleration electric fields for charges moving with constant velocities. \ On
the other hand, the magnetic field energy can also change due to accelerations
$\mathbf{a}_{e}$ and $\mathbf{a}_{q}$ of the charges. \ In this case, the
change in magnetic field energy is associated with the Faraday acceleration
terms in the electric field given in Eq. (\ref{Eqrt}).

Particularly for situations where the speed of current carriers varies, the
vector potential $\mathbf{A}\left(  \mathbf{r,}t\right)  $ serves as an
intermediate connection between magnetic fields and electric fields through
$\mathbf{B=}\nabla\times\mathbf{A}$ and $\mathbf{E=-}\nabla\Phi\mathbf{-(}%
1/c\mathbf{)\partial A/\partial}t$. \ Any external or internal agent that
tries to accelerate the current carriers for a magnetic field produces a
time-changing vector potential, which creates an electric field, that (fitting
with Lenz's law) tries to balance the energy change against work done by the
external or internal agent. \ The change in the speed of the current carriers
may produce a change in the \textit{magnetic} field energy, and this change in
\textit{magnetic} field energy is consistent with energy conservation because
the external or internal agent causing the change had to do additional work
against the back Faraday \textit{electric} fields of the accelerating charges.

\section{Acknowledgement}

This article was stimulated by a manuscript of Dr. Hanno Ess\'{e}n discussing
magnetic energy changes, \textquotedblleft A classical Aharonov-Bohm effect
arises when one goes beyond the test particle approximation.\textquotedblright%
\ \ I am grateful to two anonymous referees whose extensive and helpful
comments improved this article.


\begin{thebibliography}{99}                                                                                               %


\bibitem {GriffithsA}D. J. Griffiths, \textit{Introduction to Electrodynamics}
4th edn (Pearson, New York 2013), pp. 247-248.

\bibitem {Shadowitz}A. Shadowitz, \textit{The Electromagnetic Field} (Dover,
New York, 1988), pp. 197, 208-209, 517-522. \ A. Garg, \textit{Classical
Electromagnetism in a Nutshell} (Princeton U. Press, Princeton, NJ 08450,
2012), pp. 107-108.

\bibitem {quantum}See for example, D. J. Griffiths, \textit{Introduction to
Quantum Mechanics }2nd ed. (Pearson Prentice Hall, Upper Saddle River, NJ
2005), pp. 384-391 or L. E. Balentine, \textit{Quantum Mechanics} (Prentice
Hall, Englewood Cliffs, New Jersey 07632, 1990), pp. 220-223.

\bibitem {AB}T. H. Boyer, \textquotedblleft The Aharonov-Bohm phase shift as a
Relativity Paradox,\textquotedblright\ and \textquotedblleft A Classical
Electromagnetic Basis for the Aharonov-Bohm Phase Shift,\textquotedblright%
\ submitted for publication.

\bibitem {Darwin}C. G. Darwin, \textquotedblleft The Dynamical Motions of
Charged Particles,\textquotedblright\ Phil. Mag. \textbf{39}, 537-551 (1920). \ 

\bibitem {Jackson593}J. D. Jackson, \textit{Classical Electrodynamics }2nd edn
(John Wiley \& Sons, New York, 1975), pp. 593-595, and p. 616, Problem 12.12.

\bibitem {PA}L. Page and N. I. Adams, \textit{Electrodynamics }(Van Nostrand,
New York, 1940), p. 175. \ See also, L. Page and N. I. Adams,
\textquotedblleft Action and reaction between moving charges," Am. J. Phys.
\textbf{13}, 141--147 (1945).

\bibitem {B1971}T. H. Boyer, \textquotedblleft Energy and Momentum in
Electromagnetic Field for Charged Particles Moving with Constant
Velocities,\textquotedblright\ Am. J. Phys. \textbf{39}, 257-270 (1971).

\bibitem {Griffiths}See ref. 1, pp. 373-378.

\bibitem {J697}See reference 6, p. 697, Problems 14.12 and 14.13.

\bibitem {lagg}T. H. Boyer, \textquotedblleft Classical electromagnetic
deflections and lag effects associated with quantum interference pattern
shifts: considerations related to the Aharonov-Bohm effect,\textquotedblright%
\ Phys. Rev. D \textbf{8}, 1679-1693 (1973); \textquotedblleft The
Aharonov-Bohm effect as a classical electromagnetic lag effect: an
electrostatic analogue and possible experimental test,\textquotedblright\ Il
Nuovo Cimento \textbf{100}, 685-701 (1987); \textquotedblleft Does the
Aharonov-Bohm effect exist?\textquotedblright\ Found. Phys. \textbf{30},
893-905 (2000); \textquotedblleft Classical electromagnetism and the
Aharonov-Bohm phase shift,\textquotedblright\ Found. Phys. \textbf{30},
907-932 (2000); \textquotedblleft Darwin-Lagrangian analysis for the
interaction of a point charge and a magnet: Considerations related to the
controversy regarding the Aharonov-Bohm and Aharonov-Casher phase
shifts,\textquotedblright\ J. Phys. A: Math. Gen. \textbf{39}, 3455-3477 (2006).

\bibitem {self}T. H. Boyer, \textquotedblleft Faraday induction and the
current carriers in a circuit,\textquotedblright\ Am. J. Phys. \textbf{83},
263-271 (2015).

\bibitem {Emc2}T. H. Boyer, \textquotedblleft Electrostatic Potential Energy
Leading to an Inertial Mass Change for a System of Two Point
Charges,\textquotedblright\ Am. J. Phys. \textbf{46}, 383-385 (1978).
\end{thebibliography}
\end{document}